\documentclass[preprint,showpacs,preprintnumbers,amsmath,amssymb,superscriptaddress]{revtex4}

\usepackage{graphicx}
\usepackage{bm}

\begin{document}

\title{Geometry-controlled  kinetics}

\author{O. B\'enichou}
\affiliation{UPMC Univ Paris 06, CNRS-UMR 7600 Laboratoire de Physique Th\'eorique de la Mati\`ere Condens\'ee,
 4 Place Jussieu, F-75005 Paris, France.}

\author{C. Chevalier}
\affiliation{UPMC Univ Paris 06, CNRS-UMR 7600 Laboratoire de Physique Th\'eorique de la Mati\`ere Condens\'ee,
 4 Place Jussieu, F-75005 Paris, France.}

\author{J. Klafter}
\affiliation{School of Chemistry, Raymond and Beverly Sackler Faculty of Exact Sciences, Tel Aviv University, Tel Aviv, Israel; and Institute for Advanced
Studies (FRIAS), University of Freiburg, 79104, Freiburg, Germany.}

\author{B. Meyer}
\affiliation{UPMC Univ Paris 06, CNRS-UMR 7600 Laboratoire de Physique Th\'eorique de la Mati\`ere Condens\'ee,
 4 Place Jussieu, F-75005 Paris, France.}

\author{R. Voituriez}
\affiliation{UPMC Univ Paris 06, CNRS-UMR 7600 Laboratoire de Physique Th\'eorique de la Mati\`ere Condens\'ee,
 4 Place Jussieu, F-75005 Paris, France.}

\date{\today}

\begin{abstract}

It has long been appreciated that transport properties can control reaction kinetics. This 
effect can be characterized by the time it takes a diffusing molecule to reach a target  --  the first-passage time (FPT). Although essential to quantify the kinetics of reactions on all time scales, 
determining the FPT distribution was deemed so far intractable. Here, 
we calculate analytically this FPT distribution and show that transport processes as various as regular diffusion, anomalous diffusion, diffusion in disordered media 
and in fractals fall into the same universality classes.  
Beyond this theoretical aspect, this result changes the views on standard reaction 
kinetics. More precisely, we argue that geometry can become a key parameter so far 
ignored in this context, and introduce the concept of "geometry-controlled kinetics". 
These findings could help understand the crucial role of spatial organization of 
genes in transcription kinetics, and more generally the impact of geometry on diffusion-limited reactions.

\end{abstract}

\maketitle

It is known on general grounds that reaction kinetics can be influenced by the transport properties of the reactants \cite{Rice:1985,Kopelman:1988,Shlesinger:1993}. Actually, the transport step, before reactants meet and eventually react, even becomes limiting in the case of  confined systems such as cells or cell subdomains
where a small number of reactants are involved \cite{Yu:2006,Elf:2007,Schuss:2007,Loverdo:2008,Mirny:2008,PCCP}. In such systems, the first step in estimating the kinetics of reactions  consists in evaluating the properties of  first-encounter  between reactants.
  Quantitatively, this amounts to calculating the distribution of the time it takes a diffusing molecule to reach a target site -- the first-passage time (FPT) distribution. While this quantity is  well-known in  quasi-one dimensional or unconfined geometries (see \cite{Redner:2001a} for a review), 
  determining the FPT distribution appears to be  intractable in the realistic situation where the diffusing molecule is confined within a finite  domain \cite{Shlesinger:2007vf}. A first estimate of the effect of geometrical parameters of confinement on this search time is given by the mean of the FPT. This has recently been calculated, and a  linear scaling with the volume has been demonstrated \cite{Condamin:2005db,Condamin:2007zl,Benichou:2008,Benichou:2008a}. However, 
  as soon as several time scales are involved, the kinetics can not be determined by  the mean of the FPT only,  and the entire distribution is needed \cite{Bouchaud:1990b,Kurzynski:1998}.

Gene transcription provides an extremely important example --  which we shall repeatedly invoke in what follows --  of  reactions  involving a small copy number  of (or even single)  reactants confined within a small (micron-sized) domain, and whose kinetics must be precisely regulated to fulfill vital cell functions.  
Interest in the question of how geometrical parameters impact on the kinetics of such transcriptional reactions and could act as regulatory factors has recently increased, mainly because of new experimental tools  that enable the observation of the real-time production of proteins at the single molecule level  \cite{Yu:2006}.  These technics, which  give access to the spatial organization of the genetic material,  
 have revealed strong correlations between the spatial locations of successively 
activated genes, both for prokaryotes \cite{Kolesov:2007a} and eukaryotes \cite{Fraser:2007}. Indeed, it has been found that successively 
activated genes are often colocalized, that is located  in the very same nuclear regions. These observations  raise the question of the importance of geometrical parameters in transcription kinetics, which has remained so far widely unanswered.

In the broader context of chemical reactions in confinement, we are  interested here in the following questions :  (i) How does the FPT distribution depend on the volume of the confining domain?  (ii) How does it depend on the initial position of the diffusing molecule, and (iii) is this geometric dependence 
 an important factor, which could potentially  control the kinetics ?
 
 Note that the influence of confinement and crowding  effects on  biochemical reactions  
has already been studied (see \cite{Zimmerman:1993,Zhou:2008} for reviews)  on the basis of 
a thermodynamical treatment of reaction kinetics.  While this approach is well suited to the case of a large number of reactants, it does not provide
  the dependence
 of  the kinetics on the geometrical parameters  mentioned above  (such as the initial position of the reactants), which involve  the individual nature of the reactants 
 and their dynamical properties.

\begin{figure}[h!]
 \includegraphics[width=0.8\linewidth,clip]{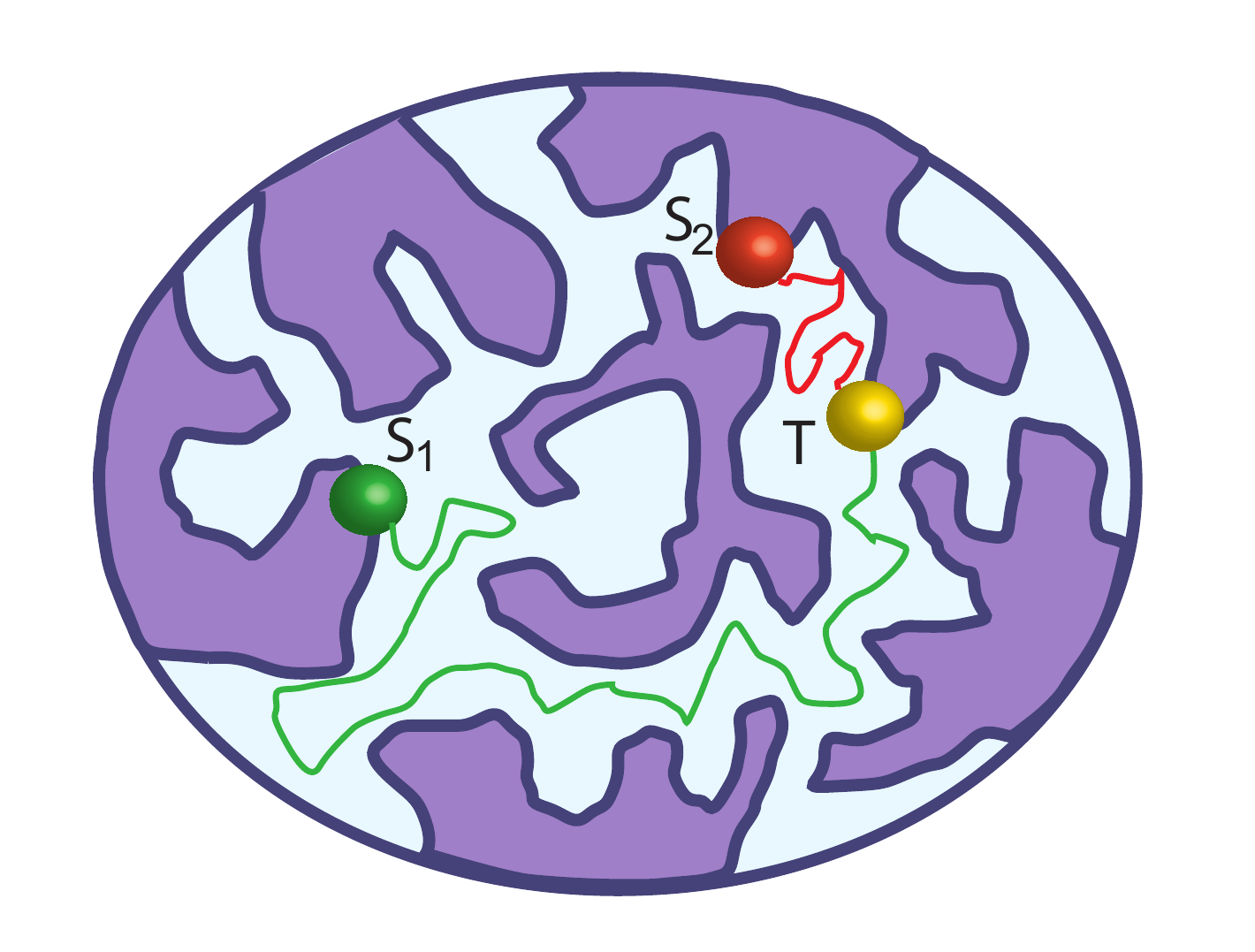}   
  \caption{}
\label{nucleus}
\end{figure}

  In this work, we calculate analytically the  distribution of the FPT at a target $T$ for a diffusing 
 particle released at a starting point $S$ (see figure \ref{nucleus}) and quantitatively answer questions (i)-(iii) above. We highlight universal laws of the FPT distribution as function of  
 the volume $N$ of the confining domain and of the distance  $ST\equiv r$ and show that two regimes emerge. More precisely, we find that the key criterion is the compact {\it vs} non compact nature of the diffusion process, to be defined mathematically below. In the non compact case, which physically corresponds to a diffusing molecule which "sparsely" explores its environment and leaves unvisited regions (typically a molecule  diffusing in a dilute solution), we show that the kinetics is widely independent of the starting point. On the opposite  compact case  of  a diffusing molecule which "densely" explores its environment   (for example a molecule in a very crowded medium), the position of the starting point strongly influences  the search
 time of the target,  which leads us to introduce the concept of "geometry--controlled kinetics". 
  In the context of gene transcription, this result implies that the kinetics of activation of a gene $T$ by a transcription factor (TF) can be orders of magnitudes faster if
 the TF is released from a site  which is colocalized with ({\it i.e.} in the vicinity of) $T$, as compared to the case where the TF is released from a remote site.

  \vspace{1cm}

{\bf Results}

We consider a Markovian random walker of position ${\bf r}(t)$, whose dynamics  is characterized by the dimension of the walk $d_w$, defined by the scaling of the mean squared displacement   $\langle {\bf r}^2(t)\rangle \propto t^{2/d_w}$. The walker is confined in a domain of $N$  with reflecting walls. Additionally, we assume that the medium is of fractal type, so that its characteristic linear size $R$ scales as  $R\propto N^{1/d_f}$, where $d_f$ is the fractal dimension \cite{D.Ben-Avraham:2000}. We are interested in the distribution $P({\bf T}_{TS})$ of the time it takes a walker   starting from the site $S$ to reach for the first time the target site $T$ located at a distance $r$ from $S$.

We start from the backward equation satisfied by the probability $P({\bf T}_{TS}=t)$ in discrete time $t$ \cite{Redner:2001a}
\begin{equation}
P({\bf T}_{TS}=t)=\sum_j w_{jS}P({\bf T}_{Tj}=t-1),
\end{equation}
 obtained by partitioning over the first step of the walk, where $w_{ij}$ stands for the transition probability from site $j$ to site $i$. It is shown in Supplementary Information (SI) that  this equation, after Laplace transform, leads  to
 the following hierarchy    satisfied by the moments $ \langle {\bf T}_{Tj}^n\rangle$ of the FPT :

\begin{equation}\label{momentsexpl}
 \langle {\bf T}_{TS}^n\rangle=\frac{1}{W_T^{\rm stat}}\sum _j \sum_{k=1}^n \binom{n}{k}(-1)^{k+1} [(H_{TT}-H_{TS})W_j^{\rm stat}+(H_{jS}-H_{jT})W_T^{\rm stat}] \langle {\bf T}_{Tj}^{n-k}\rangle.
\end{equation}
Here
\begin{equation}
 H_{ji}=\sum_{n=0}^\infty(W_{ji}(n)-W_{j}^{\rm stat})
\end{equation}
and $W_{ji}(n)$ denotes the propagator of the walk, ie the probability to be at site $j$ at step $n$ starting from site $i$, while $W_{j}^{\rm stat}$ is the probability to be at site $j$ in the stationary state.

At this stage, the hierarchy  of equations (\ref{momentsexpl}) remains formal, as it involves the unknown functions $H_{ji}$, and does not allow an explicit determination of the FPT distribution.  However,  this difficulty can be circumvented by taking the large volume limit and by considering  the rescaled time $\theta=  {\bf T}_{TS} /\overline{\langle {\bf T} \rangle}_T$, 
where $\overline{\langle {\bf T} \rangle}_T=\sum_S \langle{\bf T}_{TS}\rangle W_{S}^{\rm stat}$ stands for the mean FPT to the target site ${\bf r}_T$, 
averaged over the initial position. Note that here  we implicitly assume that $\overline{\langle {\bf T} \rangle}_T$ (as well as its disorder average in the case of disordered systems to be discussed below) is finite.
Actually, a detailed analysis of Eq.(\ref{momentsexpl}) allows us to show in SI that the distribution of  the rescaled variable $\theta$ takes in the large volume limit the following general form
\begin{equation}\label{generic}
G_{TS}(\theta)=(1-\Pi_{TS}) \delta(\theta) + \Pi _{TS}\psi(\theta)
\end{equation}
where the Dirac $\delta$ function corresponds to trajectories hitting the target without reaching the boundary  within a time of order $r^{d_w}$ much smaller than $\overline{\langle {\bf T} \rangle}_T$. The geometrical factor $1-\Pi_{TS}$
can be interpreted as  the weight of these trajectories. 
Similarly, the contribution $\Pi _{TS}\psi(\theta)$ accounts for trajectories reaching the boundary before the target. Note that the dependence on the starting point lies entirely in the geometrical factor $\Pi_{TS}$, whereas the time dependence is contained in
the scaling function $\psi$. 
The geometrical factor $ \Pi _{TS}$ and  the rescaled variable $\theta$ are explicitly determined in SI  and  their scaling with the volume $N$ and the distance $r$ are obtained under the standard scale-invariance assumption of the unconfined propagator 
$W_{ij}^{\infty}(n)\propto n^{-d_f/d_w}f \left(|{\bf r}_i- {\bf r}_j|/n^{1/d_w}\right)$ \cite{D.Ben-Avraham:2000}.   
Actually, the FPT distribution given by Eq.  (\ref{generic}) falls into a few universality classes, defined  according to  a purely geometrical criterion  as detailed below.
 
In  the case of non compact exploration defined here by $d_w<d_f$    \cite{Gennes:1982}, where the mean number of distinct sites visited by the walker in absence of confinement grows linearly with the number of steps, we find :
\begin{equation}
\label{noncompact}
\left\{
\begin{array}{lll}\displaystyle
     \overline{\langle {\bf T} \rangle}_T = H_{TT}/W_T^{\rm stat} \propto N  \\\displaystyle
    \Pi_{TS} = \frac{\langle {\bf T}_{TS}\rangle}{\overline{\langle {\bf T} \rangle}_T} \propto 1-\alpha\left(\frac{1}{r}\right)^{d_f-d_w}\\\displaystyle
     \psi(\theta)=e^{-\theta}\displaystyle
\end{array}
\right.
\end{equation}
where $\alpha$ is a lattice dependent constant of order 1. Note that the linear dependence on  $N$ of the scaling variable $\overline{\langle {\bf T} \rangle}_T $  is the same as found in  \cite{Condamin:2007zl} for the mean FPT
$\langle {\bf T}_{TS}\rangle$, and in particular does not depend of the dimensions $d_f$ and $d_w$. This result includes as a special case regular diffusion in 3D given in   \cite{Condamin:2007eu} for
which $d_w=2$ and $d_f=3$. Strikingly, the exponential  form of $\psi$ 
holds for any dimensions such that $d_w<d_f$.
Note that while the FPT distribution is a mere exponential of weight one  in the limit of $r$ larger than the step size, it departs significantly from this distribution if the starting point 
is close to the target.

In the opposite case of compact exploration $d_w>d_f$ \cite{Gennes:1982}, where the mean number of distinct sites visited by the walker in absence of confinement grows slower than linearly with the number of steps,   further hypothesis on the unconfined propagator are needed to estimate the relative importance of the terms involved in Eq. (\ref{momentsexpl}). Making use of  the O'Shaughnessy-Procaccia operator \cite{OShaughnessy:1985a} to evaluate the large volume behaviour of $H_{ij}$ in (\ref{momentsexpl}), we find that the FPT distribution obeys the generic form of Eq. (\ref{generic}) with
\begin{equation}
\label{compact}
\left\{
\begin{array}{lll}\displaystyle
     \overline{\langle {\bf T} \rangle}_T = H_{TT}/W_T^{\rm stat} \propto N^{d_w/d_f}\\ \displaystyle
    \Pi_{TS} = \frac{2d_f^2}{d_w(d_f+d_w)}  \frac{\langle {\bf T}_{TS}\rangle}{\overline{\langle {\bf T} \rangle}_T}
     \propto   \left(\frac{r}{R}\right)^{d_w-d_f} \\ \displaystyle
     \psi(\theta)=\frac{2d_f d_w}{d_w^2-d_f^2}\frac{\Gamma(\nu)}{\Gamma(2-\nu)}\sum_{k=0}^\infty \frac{(\frac{\alpha_k}{2})^{3-2\nu} J_\nu(\alpha_k)}{J_{1-\nu}(\alpha_k)}e^{-\frac{\alpha_k^2 d_w d_f}{2 (d_w^2-d_f^2)}
     \theta} 
      \displaystyle
\end{array}
\right.
\end{equation}
where $\theta>0$,   $\nu=d_f/d_w$ and  $\alpha_0<\alpha_1<...$ stand for the zeros of the Bessel function $J_{-\nu}$. Strikingly, the scaling with $N$ of the scaling variable $\theta$ is 
no longer given by the mean FPT
$\langle {\bf T}_{TS}\rangle$, which clearly indicates that several time scales are involved in the problem. The interplay between these time scales leads to a non trivial family of universal non exponential scaling functions, parametrized by $d_w$ and $d_f$. The geometrical factor strongly depends on the source-target distance $r$, and in particular is  very small for $r\ll R$, as opposed to the non compact situation.
We 	add  that the marginal case $d_w=d_f$, which is compact according to the definition given in  \cite{Gennes:1982},  corresponds  to an exponential scaling function $\psi$  as given by Eq. (\ref{noncompact}), with  logarithmic 
corrections in the scalings 
of   $ \overline{\langle {\bf T} \rangle}_T$ and $ \Pi_{TS}$ with $r$ and $N$ (see SI).

Eqs(\ref{noncompact})-(\ref{compact}) fully define universality classes of FPT distributions in confinement.  Additional comments are in order: (i) Whereas the linear scaling with $N$ of the first moment is universal, the
scaling of higher moments differ in compact and non compact cases. In particular, this scaling implies that while
the reduced variance of the FPT is always of order one in the non compact case, in the compact case it reads $(\langle {\bf T}_{TS}^2\rangle-\langle {\bf T}_{TS}\rangle^2)/\langle {\bf T}_{TS}\rangle^2 \propto (R/r)^{d_w-d_f}$, so 
very large fluctuations occur for $r\ll R$. (ii) Remarkably, the FPT distribution is entirely determined as soon as the mean $\langle {\bf T}_{TS}\rangle$ of the FPT is known (as well as its average
over the starting point $\overline{\langle {\bf T} \rangle}_T$ \cite{Montroll:1969a,Kozak:2002,Agliari:2008,Haynes:2008,Tejedor:2009rc}), even if the distribution is not exponential. (iii) For specific cases, 
this mean FPT can be calculated exactly, which provides a fully explicit 
expression of the FPT distribution  (as used in  fig 2.b, 3.a, 3.b) (iv) In all cases it can be 
calculated in the large volume limit  using the recent result \cite{Condamin:2007zl}, leading to the scaling in the geometrical parameters given in Eqs(\ref{noncompact})-(\ref{compact}) .

We note that our approach covers in particular the important case of subdiffusion \cite{R.Metzler:2000}, which is characterized by a sublinear dependence of the mean squared displacement with time (that is $d_w>2$). Subdiffusion is  widespread in complex crowded environments such as biological cells \cite{Wachsmuth2000,Golding:2006}, and might  physically 
originate  from a few
classes of  models based on different underlying microscopic mechanisms \cite{Condamin:2008}.   Importantly, subdiffusive processes can be either compact or non-compact, which will prove below  to be the relevant
criterion in the context of reaction kinetics. The FPT distribution for one class of 
models for subdiffusion, which  rely on spatial inhomogeneities \cite{D.Ben-Avraham:2000} as exemplified by diffusion in  fractals,   is directly given by Eqs.(\ref{noncompact})-(\ref{compact}). Another class 
of models  stems from large trapping times, leading to the case of infinite  $\overline{\langle {\bf T} \rangle}_T$, which we have  discarded so far. While the quenched version of this type of model becomes quite involved in the case of broad distribution of trapping times  over the disorder, the FPT distribution in the annealed 
case -- the continuous time random walk model (CTRW),  which is  a standard random walk with random waiting times, 
drawn from a distribution $f(t)$ \cite{Montroll:1965,Shlesinger:1974,R.Metzler:2000,Condamin:2007yg} --  is  straightforwardly deduced  from Eqs.(\ref{noncompact})-(\ref{compact}). In this case,  the 
Laplace transform of the FPT distribution reads: $\widehat{P}_{\rm{CTRW}}(s)=\widehat{P}(\widehat{f}(s))$, where
$\widehat{P}(s)$ is the generating function (discrete Laplace transform) of the FPT distribution of the underlying discrete time random walk, which   is determined in Eqs.(\ref{noncompact})-(\ref{compact}).

\begin{figure}[h!]
  \includegraphics[width=15cm]{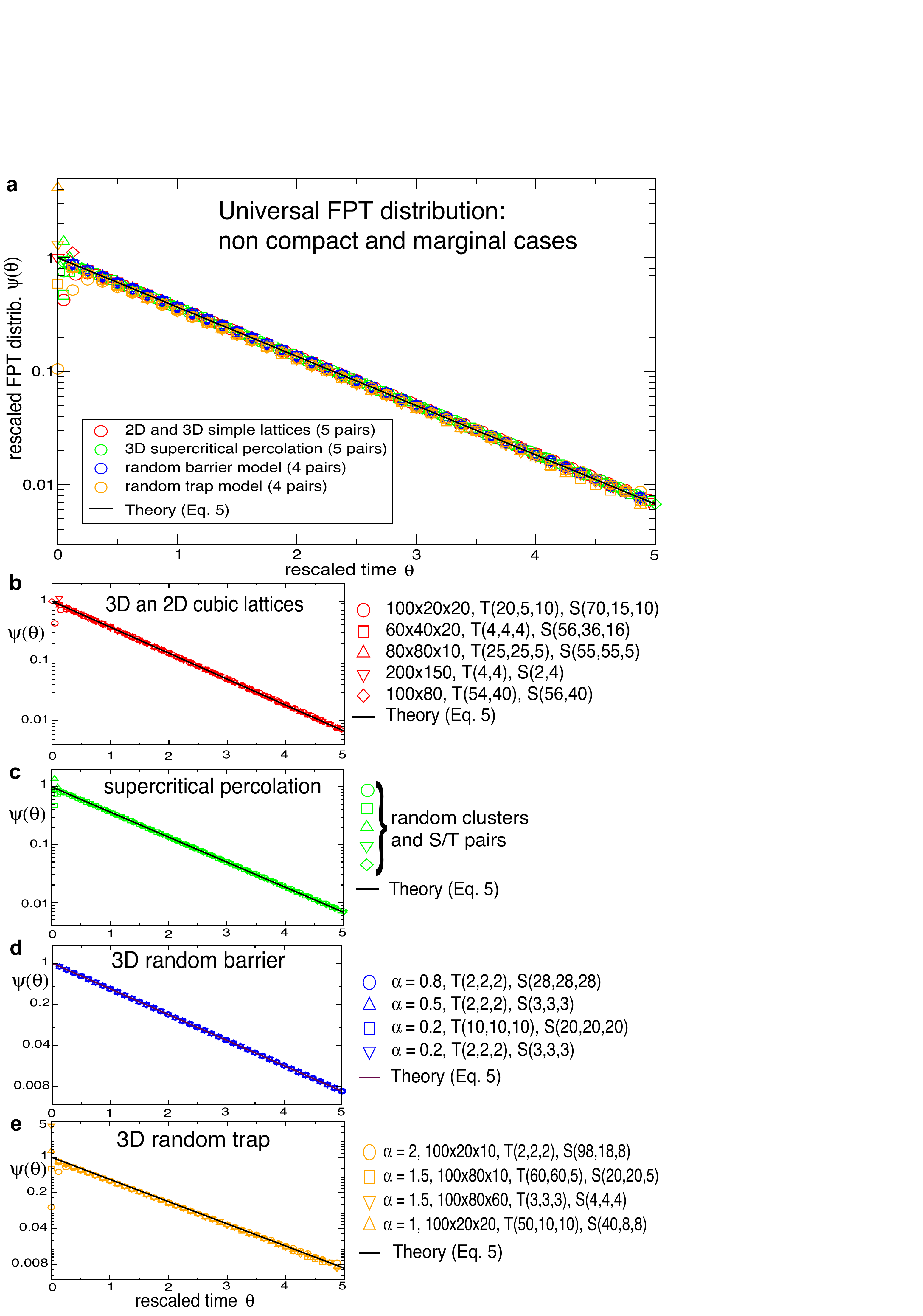}   

  \caption{}
 \label{fignoncompact}
\end{figure}

\begin{figure}[h!]
 \includegraphics[width=15cm]{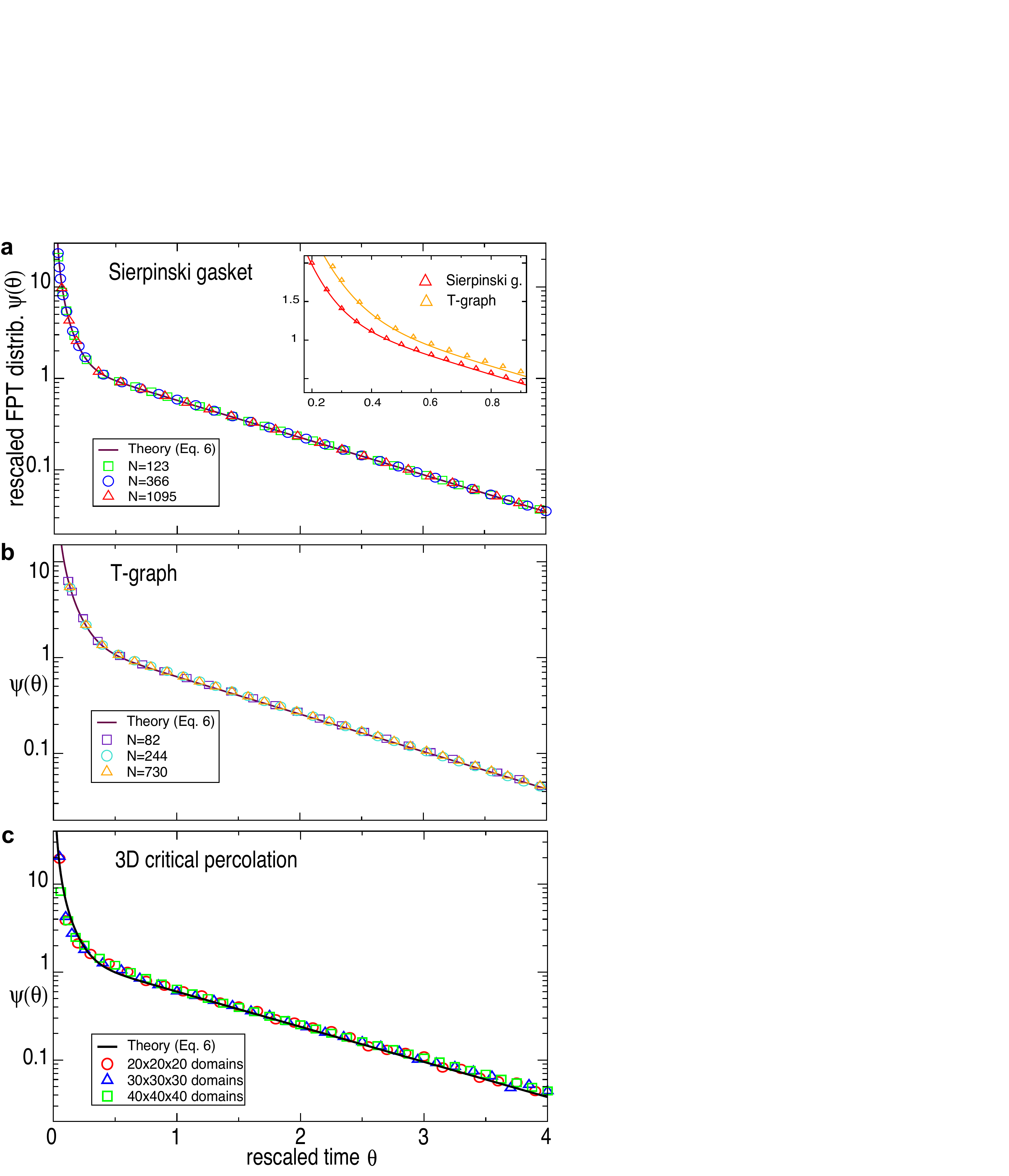}   
\caption{}
\label{figcompact}

\end{figure}

These analytical results are validated by Monte Carlo simulations and exact enumeration methods applied to various models which illustrate  the universality classes defined above. These schematic models have been widely used to describe 
transport  in disordered media \cite{D.Ben-Avraham:2000,Bouchaud:1990b}, for example in the case of exciton trapping on percolation systems \cite{Parson:1982} or  anomalous diffusion in biological cells  \cite{Saxton:2008,Malchus}, as a first step to account 
for   geometrical obstruction and binding effects involved in  real crowded environments \cite{Zimmerman:1993,Zhou:2008}.
 (i) The non compact 	and marginal cases  (see fig \ref{fignoncompact}) are
exemplified by regular diffusion on a 3D and 2D cubic lattices,
diffusion on a 3D percolation cluster above criticality, and diffusion in disordered systems  such as the   random barrier model (namely a symmetric   random walk on a 3D cubic lattice with transition rates $\Gamma$ drawn from the normalized  distribution $\rho(\Gamma)\propto \Gamma^{-\alpha}$) and the random trap model (namely a symmetric   random walk on a 3D cubic lattice with frozen waiting times $\tau_i$ at each site drawn from the normalized  distribution $\rho(\tau)\propto \tau^{-(1+\alpha)}$) . (ii) The compact case (see fig \ref{figcompact}) is exemplified by   diffusion on deterministic fractals such as
Sierpinski gasket and T-graph  and on a critical percolation cluster, as defined in fig 1 of the SI. Figures \ref{fignoncompact} and \ref{figcompact} reveal an excellent quantitative agreement between the asymptotic analytical predictions and the numerical simulations,  even for systems of small size. We emphasize that despite their very different nature, all these
models fall into the above defined universality classes.

\vspace{1cm}

{\bf Discussion}

\begin{figure}[h!]

\includegraphics[width=18cm]{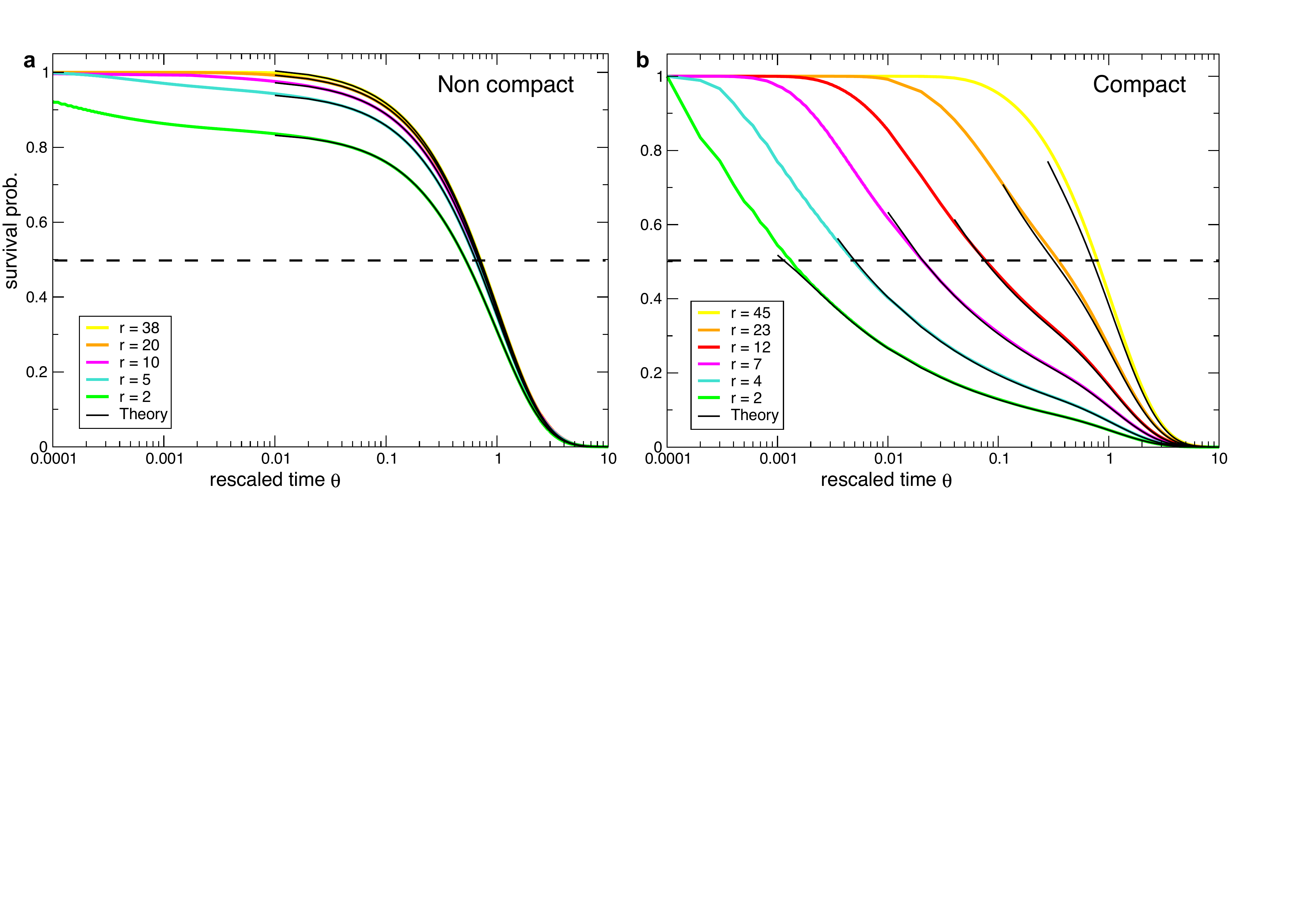}

\caption{}\label{figsurvie}
\end{figure}

We now discuss important implications of these results for reaction kinetics, using the ubiquitous  example of a target search process involving an immobile target $B$ and  searcher particles $A$ \cite{Rice:1985}. 
 When only a small number of $A$ are involved in the reaction, as is the case in a microdomain in a biological cell, this reaction has to be described at the single molecule level \cite{Schuss:2007} and  can  be quantified by the  survival probability of a particle $A$, $\displaystyle S(t)=1-\sum_{t'=0}^tP(t')$, which gives the probability that $A$ has not reacted with $B$  until time $t$. The 
quantity $\displaystyle S(t)$  depends on the initial position of the molecule $A$ and is explicitly determined using Eqs (\ref{noncompact})-(\ref{compact}), which indicate that such reactions in confinement obey  universal kinetic laws,
depending on the non compact {\it vs} compact nature of the underlying transport process:

(i) In the non compact case, which corresponds qualitatively to trajectories leaving  many sites unvisited, as in the case  of a 3D medium dilute enough to lead to regular diffusion, for any $r$ 
significantly greater than a typical  molecular length, the geometrical factor $\Pi_{TS}$ is close to 1, and the dependence on the initial position is lost. We therefore recover a simple  first order decay of the survival probability which 
 depends on the volume of the confining domain only, and not on the initial position of the reactant.  In this case of non compact exploration, 
the initial position is not an important parameter of the kinetics (see fig. \ref{figsurvie}), except in specific cases involving return times,  such as recombination reactions .

(ii) On the contrary,  in the compact case, which means
 that   each visited site is on average oversampled, geometrical factors dominate. This is typically  the case of a crowded medium described to a first approximation as a fractal structure where  the available space for diffusion is  restricted. Here,  the temporal evolution of  $S(t)$ strongly depends on the starting position. $S(t)$ drops to small values, indicating that the reaction  occurred with high probability, on a time scale which depends on the volume, but also critically  on the starting position of the reactant. This time scale ranges according to  Eq(\ref{generic}), (\ref{compact})
from $r^{d_w}$ (for starting positions such that $r\ll R$) to $R^{d_w}$ (for starting positions such that $r\simeq R$) , which in practice can span several orders of magnitude (see fig. \ref{figsurvie}).  
 In this type of "geometry-controlled reactions" (not to be confused with "fractal-like reactions"  \cite{Kopelman:1988}), spatial organization of reactants plays a crucial role, which can be quantified by our approach. 
  
 We stress that the decisive criterion leading to geometry controlled kinetics is not the subdiffusive {\it vs} diffusive nature of the transport process, but its compact {\it vs} non compact type.
We expect this effect to impact a wide class of  reactions involving either an inhomogeneous initial concentration  of reactants,  such as a speckeled distribution as experimentally realized in \cite{Monson:2000}, or  a small number of particles, such as biochemical reactions in cell subdomains. Notably, in the context of gene colocalization,
our results  give access to the kinetics of elementary steps of  activation  by transcription factors. 
As an illustrative example, let us consider two genes $A$ and $B$ 
which 
share a common transcription factor (TF) (for example 
the genes $sog$ and $zen$ of the $Drosophila$ genome which are both targeted by Dorsal  \cite{Markstein:2002}). 
Experimental results concerning subdiffusive motion of tracer particles in the 
nucleus \cite{Bancaud:2009,Wachsmuth2000,Platani:2002}
on the one hand and observations of a fractal organization 
of the chromatin on the other hand \cite{Lebedev:2005,Lebedev:2008,Bancaud:2009}, provide the following estimates
  $2\le d_w\le 3$ and $d_f\simeq2.4$, and therefore suggest that both compact and non compact exploration cases could occur.    
Relying on the analysis of  the survival probability  developed previously, 
we find in the case of compact exploration (with for example $d_f= 2.4$,  and $d_w=3$ as in \cite{Platani:2002}) that the typical time needed 
for the TF  to reach gene $B$ starting from a  gene $A$ colocalized  with, {\it i.e.} in the vicinity of,  $B$ (typically $r_{AB}=r_{\rm coloc}\le 100  \, \mbox{nm}$, which is the size of a transcription factory \cite{Fraser:2007})
can be $(r_{\rm remote}/r_{\rm coloc})^{d_w}\simeq10^{3}$ times faster than for a remote  gene $A$ (typically $r_{AB}=r_{\rm remote}\simeq 1  \, \mbox{$\mu$ m}$, which is the order of magnitude of a nucleus radius). This is in strong contrast with the case of non compact exploration (with
 for example $d_f= 2.4$, and $d_w=2$ \cite{Wachsmuth2000}) where the typical activation time of B starting from A has the same order of magnitude for A either  colocalized with B or remote. 
In other words, this suggests that gene colocalization is highly favorable for transcription kinetics, but only when it is geometrically-controlled, that is in the case of compact exploration, which makes the experimental characterization of the nature of transport in the nucleus a major issue.

To conclude,   
we calculated analytically the FPT distribution of a diffusing particle to a  target, and showed that transport processes as various as regular diffusion, anomalous diffusion, diffusion in disordered media 
and in fractals fall into the same universality classes.  Our results put forward that geometry, and in particular the initial localization of reactants,  can become a key parameter of reaction kinetics in confinement. 
In particular,  this regime of "geometry-controlled kinetics" could be relevant to transcription kinetics and could help understand the crucial role of spatial organization of 
genes.

\vspace {1cm}

{\bf Acknowledgments} 

Support from ANR grants DYOPTRI and DYNAFT is acknowledged.

{\bf Author contributions} 

All authors contributed equally to this work.

\bibliographystyle{naturemag.bst}


\newpage

Figure captions :

Fig 1 :  {\bf First-passage time distribution (FPT) and geometry controlled kinetics}. Is (or is not) the initial position of the particle an
 important parameter of the kinetics ? We show quantitatively  that in the case of compact exploration  (eg for dilute solutions), the kinetics turns out to be widely independent of the starting point ($S_1$ or $S_2$), whereas    in the non compact exploration case (eg for  crowded environments), the position of the starting point strongly influences  the search
 time of the target, 
 leading to "geometry controlled kinetics". This result in particular implies that the kinetics of activation of a gene $T$ by a transcription factor (TF) can be orders of magnitudes faster if
 the TF is released from a site $S\equiv S_2$ which is colocalized with ({\it i.e.} in the vicinity of) $T$, as compared to the case where the TF is released from a remote site $S\equiv S_1$.

 Fig 2 : {\bf Universal FPT distribution in the non compact and marginal cases}. 
   The simulated distribution $G_{TS}(\theta)$  divided by the weight $\Pi_{TS}$ is plotted against the universal theoretical prediction $\psi(\theta)$ Eq(\ref{noncompact}). The collapse of various examples onto a single master curve shows the universality of the result. {\bf a} All non compact and marginal cases (plotted independently in {\bf  b}, {\bf c}, {\bf d}, {\bf e}), collapse onto a single exponential master curve. 
{\bf  b}:  Regular diffusion on a 3D cubic lattice  and 2D square lattice for various rectangular  domains (of sizes $L_1\times L_2 \times L_3$ and $L_1\times L_2$) and source-target pairs $S(x,y,z)$ and $T(x,y,z)$ whose rectangular coordinates are  indicated in the legend inset.  Here $\theta$ and $\Pi_{TS}$ are calculated using exact results for $\langle {\bf T}_{TS}\rangle$ and $\overline{\langle {\bf T} \rangle}_T$
 given in \cite{Condamin:2007eu}. {\bf c}, {\bf d}, {\bf e}: 
  examples of disordered systems.  Here,  $\langle {\bf T}_{TS}\rangle$ and $\overline{\langle {\bf T} \rangle}_T$  are evaluated numerically. {\bf c}: Diffusion on a 3D percolation cluster above criticality embedded in a $30\times 30 \times 30$ rectangular domain with  link probability $p=0.4$.   {\bf d}: 3D
 random barrier model (see text) embedded in a $30\times 30 \times 30$ rectangular domain.  
 {\bf e}: 3D
 random trap model (see text) embedded in various rectangular domains.

 Fig 3 : {\bf Universal FPT distributions in the  compact case}. The simulated distribution $G_{TS}(\theta)$  divided by the weight $\Pi_{TS}$ is plotted against the universal theoretical prediction $\psi(\theta)$, Eq(\ref{compact}). The collapse for different system sizes $N$ shows the universality of the results. 
 Examples of deterministic fractals: {\bf a}  Diffusion 
on a Sierpinski gasket (with a target   at the apex site) and {\bf b} on a T graph (with a target at the center).  Here,  exact expressions are used for calculating $\langle {\bf T}_{TS}\rangle$ and $\overline{\langle {\bf T} \rangle}_T$. The insert  
shows that the scaling function $\psi$ weakly depends on the dimensions $d_f$ and $d_w$. 
{\bf c} Diffusion on a 3D critical percolation cluster (random fractal) embedded in rectangular domains of sizes $(L_1\times L_2 \times L_3)$, as indicated in the inset.  Here,  $\langle {\bf T}_{TS}\rangle$ and $\overline{\langle {\bf T} \rangle}_T$  are evaluated numerically, and  average over pairs of points is performed, to fulfill the scale-invariance hypothesis of the propagator (see text before Eq. (\ref{noncompact})).

Fig 4 : {\bf Reaction kinetics as quantified by the survival probability $S(t)$}, plotted for different source-target distances $r$. The non compact case  {\bf a} is exemplified by a 3D regular diffusion and the compact case {\bf b} by a diffusion on a Sierpinski gasket (with a target at the apex). The theoretical prediction for $S(t)$ is obtained from Eqs.(\ref{noncompact})-(\ref{compact}). Here $\theta$ and $\Pi_{TS}$ are calculated using exact results for $\langle {\bf T}_{TS}\rangle$ and $\overline{\langle {\bf T} \rangle}_T$.
Quantitatively, the typical reaction time $t_{{\rm typ}}$ can be defined, for example, through the median $S(t_{{\rm typ}})=1/2$, indicated by the dotted line. In the non compact case,  $t_{{\rm typ}}$  weakly 
depends on the initial position of the reactant, which therefore is not an important parameter of the kinetics.   On the contrary, in the compact case, $t_{{\rm typ}}$  runs over several orders of magnitude  depending on 
the initial position which, in turn, controls the kinetics.

\newpage

\newpage

\newpage

\newpage

\end{document}